\documentclass[10pt,a4paper,twocolumn,superscriptaddress]{revtex4}

\usepackage{textcomp}
\usepackage{color}
\usepackage{amsmath}
\usepackage{mathrsfs}
\usepackage{graphicx}
\usepackage{subfig}
\usepackage{epsfig}
\usepackage{hyperref}

\newcommand{\microw}{\textmu{W }}

\usepackage[squaren,Gray]{SIunits}
\usepackage[final]{pdfpages}

\begin{document}

\title{Temporal Solitons in a Coherently Driven Active Resonator}

\author{Nicolas Englebert}
\email{nicolas.englebert@ulb.ac.be}
\author{Carlos Mas Arabí}
\author{Pedro Parra-Rivas}
\author{Simon-Pierre Gorza}
\author{Fran\c{c}ois Leo}

\affiliation{Service OPERA-\textit{Photonique}, Universit\'e libre de Bruxelles (U.L.B.), 50~Avenue F. D. Roosevelt, CP 194/5, B-1050 Brussels, Belgium}

\begin{abstract}
Optical frequency combs are lightwaves composed of a large number of equidistant spectral lines. They are important for metrology, spectroscopy, communications and fundamental science. Frequency combs are most often generated by exciting dissipative solitons in lasers or in passive resonators, both of which suffer from significant limitations.
Here, we show that the advantages of each platform can be combined.
We introduce a novel kind of soliton, called active cavity soliton, hosted in coherently driven lasers pumped below the lasing threshold. We use an active fibre resonator and measure high peak power solitons on a low power background, in excellent agreement with simulations of a generalized Lugiato-Lefever equation.
Moreover, we find that amplified spontaneous emission has negligible impact on the soliton’s stability. Our results open up novel avenues for frequency comb formation by showing that coherent driving and incoherent pumping can be efficiently combined to generate a high-power ultra-stable pulse train.
\end{abstract}

\maketitle

Temporal dissipative solitons are optical pulses propagating indefinitely in a nonlinear optical resonator~\cite{akhmediev_dissipative_2005}. They are sustained by a double balance of gain and loss on the one hand, and nonlinearity and dispersion on the other. 
Dissipative solitons can take many forms~\cite{grelu_dissipative_2012} and emerge in various optical systems~\cite{akhmediev_dissipative_2008}. In particular, the sech-shaped solutions of mode-locked lasers~\cite{hofer_mode_1991,haus_mode-locking_2000,wang_wideband-tuneable_2008,quarterman_passively_2009,kieu_sub-100_2009,piccardo_frequency_2020} and passive Kerr resonators~\cite{wabnitz_suppression_1993,leo_temporal_2010,herr_temporal_2014,yi_soliton_2015,brasch_photonic_2016,lilienfein_temporal_2019}, are attracting a lot of attention.
As they periodically leave the cavity, they form a stable pulse train --- an optical frequency comb (OFC) --- that finds applications in many fields, most notably metrology~\cite{udem_optical_2002}.
These two classes of solitons mainly differ by their pumping scheme.
In lasers, the soliton's energy is maintained by intracavity amplification while in passive cavities, it is sustained by coherent driving. This leads to fundamental differences between the pulse trains and corresponding OFCs, in terms of energy and coherence. 
Laser solitons are susceptible to timing jitter as amplified spontaneous emission (ASE) adds random fluctuations which, in turn, causes the central frequency to drift~\cite{gordon_random_1986,kim_ultralow-noise_2016}.
CSs, on the other hand, are phase locked to a coherent driving laser~\cite{leo_temporal_2010}.
But, as they propagate in high finesse resonators, only a small fraction of the energy can be extracted. 
Several efforts to combine the advantages of coherent driving and incoherent pumping have been reported~\cite{yoshitomi_ultralow-jitter_2006,quinlan_greater_2007,rebrova_stabilization_2010,garbin_topological_2015,bao_laser_2019} but these systems, operated above the lasing threshold, are still subject to phase and amplitude noise.

Here, we introduce a novel concept for intracavity frequency conversion, which we apply to OFC generation.
We show that by combining coherent driving and intracavity amplification, but keeping the system \emph{below} the lasing threshold, robust nonlinear attractors emerge.
Specifically, we investigate soliton formation in a Kerr resonator incorporating an amplifier.
Intracavity amplification has been proposed to tune the coupling regime of driven resonators~\cite{choi_control_2001,dumeige_determination_2008} and to improve optical gyroscopes~\cite{hsiao_planar_2007}. It was also used in the first demonstration of spatial cavity solitons in a semiconductor microcavity~\cite{barland_cavity_2002}.
Our work extends the applications of intracavity amplification by showing that it enhances frequency conversion in coherently driven resonators.

\section*{Results}
\begin{figure*}
  \centerline{\includegraphics[scale=1]{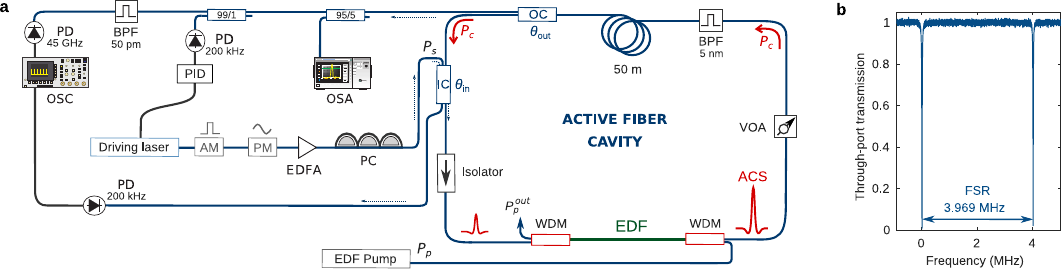}}
  \caption{\textbf{Experimental set-up and linear characterization}. \textbf{a}, Active cavity solitons (ACSs) are excited in a fibre cavity. It includes a short erbium doped fibre (EDF), pumped by a 1480~nm laser, and is externally driven by either a continuous wave, or an amplitude and phase modulated (AM and PM), 1550~nm laser.
  The driving beam is amplified by an erbium doped fibre amplifier (EDFA) and injected in the cavity through the input coupler (IC). The input polarization state is adjusted using a polarization controller (PC). The 1550~nm and 1480~nm beams are combined and separated through wavelength division multiplexers (WDM). The length of the EDF, the  1480~nm pump power, the variable optical attenuator (VOA) and the intracavity bandpass filter (BPF) are chosen so that there is no emission when the driving laser is switched off. An output coupler (OC) is included for soliton analysis, using an optical spectrum analyzer (OSA), a fast photodiode (PD) and an oscilloscope (OSC). The cavity is actively stabilized using a proportional-integral-derivative (PID) controller. \textbf{b}, Cavity resonances as measured in the through-port for $P_s=1$~\microw and $P_p=1\,$W.}
  \label{fig:setup}
\end{figure*}

The dynamics of passive Kerr resonators is well described by the seminal Lugiato-Lefever equation~\cite{lugiato_spatial_1987,haelterman_dissipative_1992}.
We introduce a generalized form that accounts for intracavity amplification (see Supplementary Information Section II). When the gain dynamics is much slower than the roundtrip time $t_R$, it reads:
\begin{equation}
\begin{array}{l}
  \hspace{-1.5mm}
  t_R\dfrac{\partial u(T,\tau)}{\partial T} = \bigg(\displaystyle -\dfrac{\Lambda}{2}+
    \dfrac{g_0L_e/2}{\displaystyle1+\left(P_{sat}t_R\right)^{-1}\displaystyle\int_0^{t_R} |u(T,\tau)|^2\,\text{d}\tau }\vspace{1mm}\\ \hspace{5mm}-i\delta_0-i\dfrac{\beta_2L}{2}\dfrac{\partial^2}{\partial \tau^2}
    +\,i\gamma L|u(T,\tau)|^2
    \bigg)u(T,\tau) + \sqrt{\theta_{\text{in}}P_s},
\end{array}
\label{eq:meanfield}
\end{equation}
where $u$ is the intracavity electric field envelope, $\Lambda$ is the intrinsic intracavity loss, $L$ is the length of the resonator, $\gamma$ its nonlinear parameter and $\beta_2$ the average group velocity dispersion.
$T$ is a slow time, defined as $T=nt_R$ where $n$ is an integer number. $\tau$ is a time reference traveling at the group velocity of the driving frequency which serves to describe the intracavity pattern.
$P_s$ is the driving power and $\delta_0$ the detuning from the closest cavity resonance. 
$g_0$ is the unsaturated gain, $L_e$ is the amplifier length and $P_{sat}$ is the gain saturation power.

Equation~\eqref{eq:meanfield} possesses many different stable nonlinear attractors. We here focus on the solitary waves. They are sech-shaped and lie on a homogeneous background. In contrast to passive Kerr resonators, these solitary waves may appear in regions without Turing patterns, nor homogeneous bistability.
This is reminiscent of laser dynamics where solitons usually do not coexist with Turing patterns~\cite{grelu_dissipative_2012}. However, the solitons of Eq.~\eqref{eq:meanfield} are phase locked to a coherent driving laser and their excitation does not require saturable absorption.
Solitons of active cavities hence constitute a novel class of localized dissipative structures. To differentiate these solitons from cavity solitons (CSs) of passive resonators and solitons of lasers, we call them active cavity solitons (ACSs). Their region of existence and stability is set by the saturation power.
In what follows we investigate, theoretically and experimentally, ACS formation in two different configurations.
We first study soliton formation under CW driving, i.e. in the regime of strong gain saturation. 
In the second part, we implement pulsed driving to decrease the average intracavity power and investigate pattern formation when gain saturation is negligible.

Our experimental demonstrations are performed with a fibre resonator.
The set-up is depicted in Fig.~\ref{fig:setup}a. The cavity is about 50~m long, mostly made of standard single mode fibre, and incorporates two couplers (90/10 and 99/1), an optical isolator and a gain section. 
The latter is mainly composed of a short erbium doped fibre pumped at 1480~nm, designed to minimize gain saturation (see Supplementary Information Section III).
The driving laser is a narrow linewidth distributed feedback laser emitting around 1550~nm. 
It can be amplitude modulated, phase modulated and/or amplified in a commercial erbium doped fibre amplifier (EDFA), before being sent to the input coupler. 
We note that each coupler can be used as an input port without any impact on the finesse. In what follows we perform experiments with both configurations depending on whether we want to extract more power from the cavity or reduce the driving power threshold. Part of the output beam is sent to a proportional-integral-derivative (PID) controller to lock the laser frequency close to a resonance of the cavity~\cite{jang_ultraweak_2013}.
\begin{figure*}
\includegraphics[scale=1]{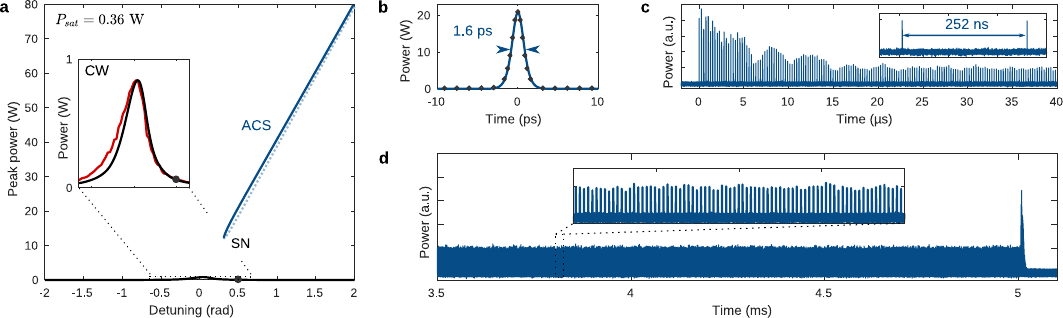}
\caption{\textbf{Soliton excitation under continuous wave driving, $\theta_{\mathrm{in}} = 10\,\%$, $P_s=110$}\,mW. {\bf a}, Stationary soliton (blue) and homogeneous (black) states of Eq.~\eqref{eq:meanfield}. The solutions are represented by their peak power. The dotted lines denote unstable attractors. SN: saddle-node bifurcation. In the inset, the theoretical homogeneous branch is compared to the experimental average output power (red) as the detuning is scanned.
The black dot indicates the stabilization setpoint ($\delta_0=0.5$). . {\bf b},\,Theoretical profile of a soliton for $\delta_0=0.5$ propagating in our active fibre cavity. The solid line corresponds to the solution of Eq.~\eqref{eq:meanfield} while the circles correspond to simulations of the full lumped model (see Supplementary Information Section I).  {\bf c},\,Oscilloscope recording when an addressing pulse is sent in the cavity. Inset: zoom on the soliton pattern, showing it is composed of short pulses separated by the roundtrip time. {\bf d}, Oscilloscope recording about 5~ms after the first addressing pulse, as a second addressing pulse is sent to erase the soliton pattern.} 
\label{fig:CW}
\end{figure*} 

We start by characterizing the resonances of the cavity.
We bypass both modulators and send the driving laser (1\,$\micro$W) to the 99/1 coupler. The 1480~nm pump power is set to 1~W.
Two cavity resonances, as measured in the through port, are shown in Fig.~\ref{fig:setup}b. The FSR is 3.969 MHz and the linewidth is 15~kHz, corresponding to a Q factor of 12.2~billion. 
Knowing the input coupler ratio, we can extract the maximum intracavity power (215\,\microw\hspace{-1.2mm}) and corresponding effective loss ($\Lambda-g_0L_e=2.4$~\%). 
When we increase the driving power, the resonances broaden because of gain saturation (see Fig.~S2a). The effective loss more than double when the intracavity power is 50~mW, and reaches 10\% around 150~mW. While these values remain low by passive fibre resonator standards
~\cite{leo_temporal_2010,jang_ultraweak_2013}, they confirm that gain saturation will play an important role in the nonlinear regime of our cavity.

To investigate soliton formation under CW driving, we set the driving power to $\theta_\mathrm{in}P_s=11$~mW where $\theta_\mathrm{in}$ is the input coupling ratio.
The corresponding theoretical resonance is shown in Fig.~\ref{fig:CW}a. The intracavity power is limited to a maximum of around 800~mW because of gain saturation.
This is below the Turing instability threshold, as confirmed by a linear stability analysis of Eq.~\eqref{eq:meanfield} (see Supplementary Information Section II).
Despite the lack of Turing instability at that driving power, numerical analysis of Eq.~\eqref{eq:meanfield} predicts ACS formation in a very large range of detunings (see Fig.\,\ref{fig:CW}a). Note that the soliton branch is very different from that of CSs, which emerge from a homogeneous saddle-node bifurcation~\cite{coen_universal_2013}. Here, the branch is disconnected from the homogeneous solution and forms an isola\,\cite{beck_snakes_2009}.
This is because the average power of the localized patterns, and hence the gain saturation, is mostly set by the level of the homogeneous background. The solitons exist in regions where the effective loss $\Lambda_e$ is low.

The ACS branch shown in Fig.\,\ref{fig:CW}a extends up to $\delta_0=4.5$ but we stress that the single resonance approximation, inherent to mean-field models, is not valid anymore in that region. 
Furthermore, the branch is limited to $\delta_0<0.8$ in our experiment. Beyond that point, the soliton disappears because of the intracavity filter which is used to suppress lasing at shorter wavelengths (see methods). 
An example of ACS, corresponding to the experimental detuning used below ($\delta_0=0.5)$, is shown in Fig.~\ref{fig:CW}b. The soliton sits atop a 60~mW homogeneous background and is 1.6 ps wide, with a peak power of 21 W. We find excellent agreement between simulations of Eq.~\eqref{eq:meanfield} and the complete lumped-element model (see Supplementary Information Section I). 

This theoretical analysis predicts the existence of novel, ultra stable and high peak power solitons in Kerr resonators incorporating an amplifier pumped below threshold. We expect the soliton branch shown in Fig.~\ref{fig:CW}a to be universal and solitons to be found in many different types of laser cavities with a strong Kerr nonlinearity.

For our experimental investigation, we use the 90/10 coupler as the input port in order to minimize the required driving power. The cavity output power, as we scan the laser frequency, is shown in Fig.~\ref{fig:CW}a. We find good agreement with the theoretical resonance. In particular, we find no evidence of modulation instability (MI).
\begin{figure*}
\includegraphics[scale=1]{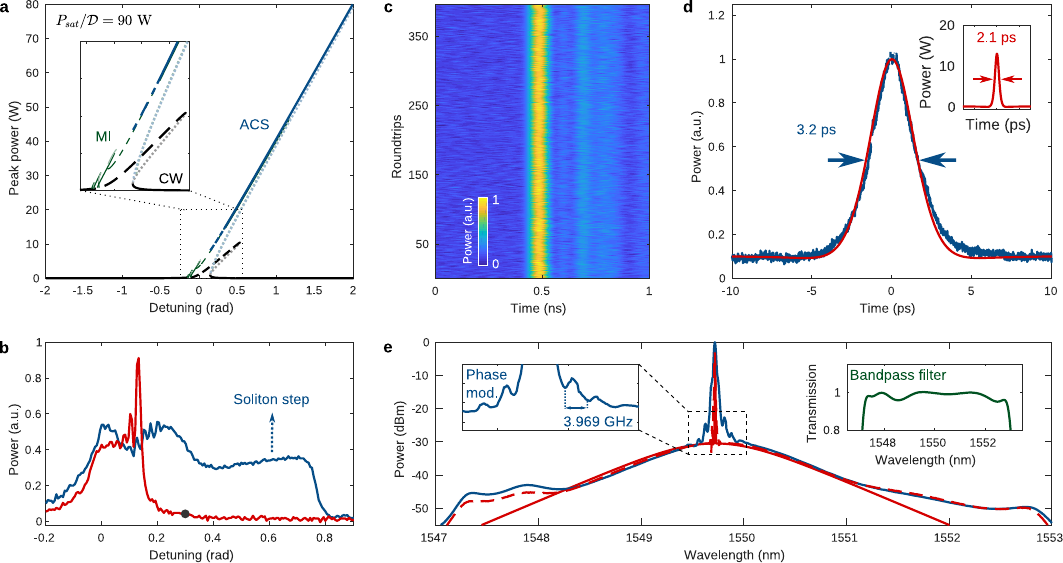}
\caption{\textbf{Soliton formation under pulsed driving, $\theta_{\mathrm{in}}=1\%$,  $\mathcal{D}=1/250$}, $P_s=750$\,mW.
{\bf a},\,Theoretical homogeneous (black), MI (green) and soliton (blue) stationary states calculated with Eq.~\eqref{eq:meanfield}. The solutions are represented by their peak power. Dotted lines correspond to homogeneously unstable states and dashed lines correspond to Hopf/MI unstable states. {\bf b},\,Experimental forward (blue line) and backward (red line) scan through a resonance. The black dot indicates the stabilization setpoint ($\delta_0=0.3$). {\bf c},\,Oscilloscope recording several seconds after the excitation process showing a stable pulse, much shorter than the 1~ns driving pulse, exiting the cavity. {\bf d},\,Experimental (blue line) and theoretical (red line) autocorrelations traces. The inset shows the theoretical profile of the corresponding soliton.
{\bf e},\,Experimental (blue) and theoretical (red) spectra at the output of the cavity. The dashed red line corresponds to the theoretical spectrum calculated with the lumped model which includes the ASE (see Supplementary Information Section I). The spectral profile of the intracavity bandpass filter, and the impact of the PM on the spectrum are both shown as insets.}
\label{fig_D_250}
\end{figure*}
The lack of Turing instability precludes the spontaneous formation of solitons in this configuration.
We hence experimentally address ACSs with a single pulse of intensity modulation~\cite{wang_addressing_2018}.
We start by stabilizing the cavity on the slope and level corresponding to $\delta_0=0.5$ (see black dot in Fig.~\ref{fig:CW}a).
We then send a 250~ps pulse with a peak power of 45~W and record the cavity output on a 10~GHz oscilloscope. 
Our results are shown in Fig.~\ref{fig:CW}. Shortly after the arrival of the addressing pulse, a pattern of solitons of constant amplitude can be seen leaving the cavity every roundtrip (Fig.~\ref{fig:CW}c).
A similar pattern is present up to the arrival time of a second addressing pulse, about 5\,ms later (Fig.~\ref{fig:CW}d), after which the solitons quickly disappear.
The delay between two addressing pulses corresponds to more than a thousand times the effective photon lifetime.
This experiment demonstrates the existence of robust and addressable short optical pulses in our cavity. Yet, the low energy level contained in a single soliton hinders a more precise temporal and spectral characterization of these ACSs in the CW-driven regime. Despite the very large peak-to-background power ratio of the soliton (see Fig.~\ref{fig:CW}b), the energy contained in the background is two orders of magnitude larger than that in the pulse.

Next, we investigate the dynamics of our resonator using flat-top pulses synchronized to the roundtrip time of the cavity. This allows examining soliton formation in a regime of lower gain saturation.
Moreover, from an experimental standpoint, it permits reaching higher driving powers and reduces the energy contained in the background, facilitating the characterization of the soliton. 
We consider a 1\,ns long driving pulse, which corresponds to a duty cycle $\mathcal{D}=1/250$, with a peak power of 750~mW. 
To facilitate comparisons between the pulsed driving regime and the CW driving regime, we introduce the effective saturation power $P_{sat}/\mathcal{D}$.
In Figure\,\ref{fig_D_250}a, we show the system's nonlinear attractors for the same pump power as above ($P_p=1$~W), which corresponds to $P_{sat}/\mathcal{D}= 90\,$W.
The branches differ from those shown in Fig.~\ref{fig:CW}a, especially close to $\delta_0=0$. In that region (shown in the inset of Fig.~\ref{fig_D_250}a), the stationary states are reminiscent of those of a passive resonator~\cite{coen_universal_2013} and similar dynamics can be expected, in particular the spontaneous formation of solitons. 
The homogeneous state is bistable and goes through a Turing instability at $\delta_0=-0.16$.
Beyond that point, several MI branches, defined as the peak power of an extended Turing pattern with a fixed period, emerge. They end up connecting to the soliton branch which no longer forms an isola. Instead, the saddle soliton emerges from a homogeneous saddle-node bifurcation. Both the soliton and MI branches go through a Hopf bifurcation~\cite{leo_dynamics_2013}.
Beyond $\delta_0=0.5$, only the soliton branch exist. In that region, it is very similar to the isola found in the CW-pumped regime (see Fig.\,\ref{fig:CW}a). The larger existence region of solitons, as compared to high average power states, is a striking dissimilarity between passive and active fibre cavities, and it persists for large effective saturation powers. 

To experimentally investigate soliton formation under pulsed driving, we carve out background-free 1~ns flat top pulses in the continuous wave driving laser with the amplitude modulator. Additionally, we imprint a 4\,GHz phase modulation on the driving beam (see Methods).
An experimental nonlinear resonance, measured by scanning the driving laser frequency, 
is shown in Fig.~\ref{fig_D_250}b. 
We observe a large bistable region, defined as the detuning range where the forward and backward scans significantly differ. 
The low power plateau, seen on the forward scan, is a common feature of nonlinear passive resonators and is often referred to as the soliton step~\cite{herr_temporal_2014,obrzud_temporal_2017}. Note that it extends up to $\delta_0=0.8$ which corresponds to the limit of our experimental setup.
These experimental scans are in good agreement with the theoretical analysis of stationary states. Turing patterns emerge close to the linear resonance and evidence of spontaneous soliton formation is found for positive detunings.
We stress that because of the dynamical nature of
a scan, corresponding to around 0.4 ms/rad in Figs.~\ref{fig_D_250}d, the measured output power is not always representative of the average power of steady state solutions at the corresponding
detuning. Yet, it indicates where to expect localized patterns in the system. 
We stabilize the cavity in that region, triggering the formation of a single soliton in the cavity (see methods).
A stable pulse train, with a period corresponding to the roundtrip time, is measured at the output coupler (see Fig.~\ref{fig_D_250}c). The pulses are resolution limited and the absence of drift indicates locking to a maximum of the $\sim4\,$GHz phase modulation. The soliton train is also sent to an autocorrelator and an optical spectrum analyzer (OSA) (see Figs.~\ref{fig_D_250}d and \ref{fig_D_250}e), confirming the excitation of a 2.1~ps pulse in the cavity.
The results are in excellent agreement with both the mean-field equation~\eqref{eq:meanfield} and the lumped-element model. The latter accounts for the intracavity filter as well as ASE (see Supplementary Information Section I).

Because the soliton sits atop a pedestal of ASE, the question arises as to whether the amplifier noise impacts the coherence of the soliton train. ASE is known to limit the coherence of dissipative solitons in lasers through the Gordon-Haus effect
~\cite{gordon_random_1986,kim_ultralow-noise_2016}. Slow spectral drifts lead to timing jitter which is detrimental to high-precision applications such as optical clocks~\cite{hinkley_atomic_2013}.
In our system, the solitons are locked to both the driving laser and a phase modulation maximum such that we expect a minimal impact of the ASE on the timing jitter. Note that we measure much less ASE when the driving pulse is longer because of the gain saturation associated with the larger average intracavity power. A spectrum from an experiment conducted with $\mathcal{D}=1/6$ is shown in the Supplementary Information (Fig.\,S3). However, in order to study the impact of noise, or lack thereof, on the stability of ACSs, we perform coherence measurements in the configuration with maximum ASE ($\mathcal{D}=1/250$).
\begin{figure}
\includegraphics[scale=1]{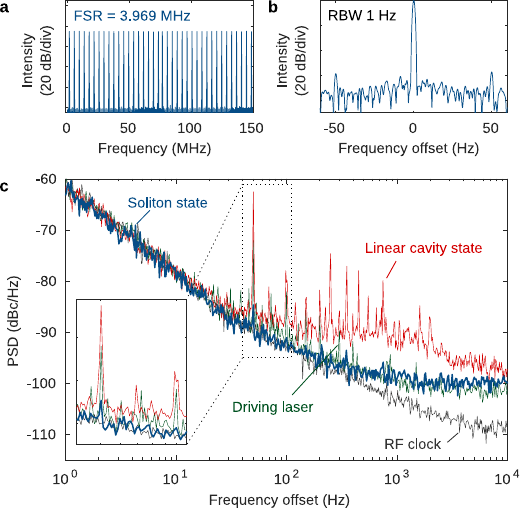}
\caption{\textbf{RF characterization of the soliton train, $\theta_{\mathrm{in}}=1\%$, $\mathcal{D}=1/250$, $P_s=750\,$}mW. {\bf a},\,RF spectrum of the soliton train showing beat notes separated by the FSR. {\bf b},\,Electrical spectrum around a single beat note (36.96~MHz). {\bf c},\,Single sideband phase noise around the 36.96~MHz beat note of the soliton (blue), the RF clock (black), the driving laser (green) as well as the linear cavity state (red). The inset highlights the noise filtering capabilities of the soliton state.}
\label{fig:RF}
\end{figure}

To experimentally characterize the stability of the pulse train, we filter out the driving laser and send the signal to a photodiode connected to an electrical spectrum analyzer (see Methods). The results are shown in Fig.~\ref{fig:RF}. The spectrum reveals a series of beat notes separated by the FSR of the cavity. The beat notes have a full width at half maximum of 1~Hz, limited by the resolution of the analyzer. We further characterize the pulse train by measuring the single sideband phase noise of a beat note around 37~MHz (see Fig.~\ref{fig:RF}c).
It is compared to the same beat note of the RF pattern, the modulated driving laser and the output of the stabilized cavity in the linear regime. The latter is obtained when the cavity is stabilized in the existence region of ACSs but none are excited ($\delta_0=0.3$, see Fig.~\ref{fig_D_250}b). 
These measurements suggest that the phase noise of the soliton train is limited by the RF clock in the low frequency range and by the noise of the driving laser at higher frequencies. Beyond 10\,kHz, and up to half the FSR, we measure a constant phase noise of around -100~dBc/Hz (not shown).
Particularly interesting is the absence of spurious peaks on the soliton phase noise, as compared to both the driving laser and the linear state, highlighting the noise filtering capabilities of ACSs. The 50~Hz peak, for example, is much less pronounced on the soliton phase noise than on the other optical measurements (see inset of Fig.~\ref{fig:RF}c). The linear cavity state is particularly noisy, possibly because of ASE, which further stresses that ACSs are robust attractors as they shed the noise present on the background.
These results prove the formation of a highly coherent soliton comb. 

\section*{Discussion}

In summary, we have shown that coherently driven active fibre resonators may host robust, addressable optical solitons which we call active cavity solitons (ACSs).
We theoretically demonstrated their wide range of existence and experimentally observed them, both with CW and pulsed driving lasers. With the former, the absence of Turing patterns prevents their spontaneous formation but we showed that they can be individually addressed. With the latter, ACSs may form spontaneously.

These solitons belong to a novel class of localized dissipative structures and combine the advantages of cavity solitons as they are highly coherent, and of laser solitons because they reshape over a single roundtrip despite the extraction of a significant fraction of their power. Our noise analysis, limited by the resolution of the equipment, revealed no impact of the ASE on the generated frequency comb. The limitations of our system, and how it compares to existing fibre combs~\cite{kim_ultralow-noise_2016}, will be the subject of future work.

Our theoretical model is universal and we expect ACSs to be found in other platforms, such as microresonators. 
The generation of CSs in passive integrated resonators (often called dissipative Kerr solitons~\cite{brasch_photonic_2016}) has attracted significant attention in the past decade. They have shown promise for important applications, including data transmission~\cite{marin-palomo_microresonator-based_2017}, optical atomic clocks~\cite{newman_architecture_2019} and microwave generation
~\cite{liu_photonic_2020}. Yet, real life applications are still lagging behind, mostly because very little power can be extracted from a high Q microresonator.
An integrated version of our hybrid cavity would combine the advantages of integrated passive resonators and lasers. Specifically, integrated resonators incorporating an amplifier, such as the ones made of doped glass~\cite{kippenberg_demonstration_2006,hsiao_planar_2007} or heterogeneously integrated III/V-on-SiN~\cite{beeck_heterogeneous_2020} are likely suitable for ACS formation as the gain can be engineered to obtain large saturation powers when pumped below threshold.

Finally, we believe that the potential applications of our hybrid cavity go beyond four-wave mixing.
Active cavities can be used to host other frequency conversion configurations. For example, the threshold of a singly resonant optical parametric oscillator~\cite{bosenberg_93_1996} would be significantly reduced by incorporating an amplifier at the signal frequency.

\section*{Methods}

\small

\subparagraph*{\hskip-10pt Bifurcation analysis}\ \\
The theoretical branches shown throughout the paper correspond to stationary solutions of Eq.~\eqref{eq:meanfield}. They are found by setting 
The stationary Turing patterns (TPs) and ACSs are calculated by means of the open distribution program AUTO-07p as boundary value problems. For the ACSs, the boundary is set to [0,$\tau_s/2$], where $\tau_s=250$\,ns in Fig.\,\ref{fig:CW}a and $\tau_s=1$\,ns in Fig.\,\ref{fig_D_250}a, and we impose Neumann boundary conditions. We calculate the stability of the stationary solutions against homogeneous perturbations by computing the  eigenvalues of the Jacobian  matrix  associated  with Eq.\,\eqref{eq:meanfield}. For the TPs, we numerically track a single pattern period ($\tau_p$) and impose periodic boundary conditions. In Fig.\,\ref{fig_D_250}a, $\tau_p$ is set to the most unstable mode at three different detunings: the MI threshold $\delta_0=-0.16$ ($\tau_p=10.3$\,ps), $\delta_0=-0.14$ ($\tau_p=10.76$\, ps)  and $\delta_0=0.01$ ($\tau_p=11.65$\,ps).

\subparagraph*{\hskip-10pt Experimental set-up}\ \\
The fibre ring cavity (see Fig.~\ref{fig:setup}a) is made of a section $L_c=50.1$\,m of standard telecommunication single-mode silica fibre (SMF-28) and a $L_e=34\,$cm section of erbium doped fibre (EDF). The latter provides the optical gain. It is surrounded by two wavelength division multiplexers (WDMs) to combine the 1480\,nm pump with the intracavity signal, and to reject the remaining pump power at the amplifier output. A backward pump is chosen to maximize the signal-to-noise ratio (SNR).
The EDF is a Liekki$^{\text{®}}$ ER16-8/125 fibre, chosen for its good spliceability with standard single mode fibres and its relatively small gain (8\,dB.m$^{-1}$ at 1550\,nm). 
Its length is empirically set so that the gain is slightly larger than the cavity loss. We then use a variable optical attenuator to increase the loss so that the cavity is below the lasing threshold. An optical bandpass filter hinders laser emission at shorter wavelengths. It has a bandwidth of 5\,nm at 0.5\,dB and is centred on 1550\,nm (its transmittance is shown in the right inset of  Fig.~\ref{fig_D_250}e). 
The cavity contains two couplers (99/1 and 90/10). They are used to either inject the driving beam into the cavity or extract part of the intracavity power. 
Finally, the dual-stage optical isolator (60\,dB of isolation) in the cavity prevents the build up of Brillouin scattering. Without the doped fibre, the total cavity loss (including the insertion loss of the two WDMs) is 31\%. The whole cavity is enclosed in a box to passively reduce the environmental perturbations. 

The driving continuous wave (CW) laser is a Koheras Adjustik$^{\text{TM}}$ E15 with a sub-100~Hz linewidth. Its wavelength is set to 1549.72\,nm, on the edge of the laser tuning range (1\,nm) to be as close as possible to a local transmission peak of the bandpass filter (1549.45\,nm, see Fig.~\ref{fig_D_250}e).  To synchronously drive the cavity, the laser output is modulated with a Mach-Zehnder amplitude modulator (bandwidth: 12 GHz, extinction ratio: 30\,dB), driven by a pattern generator
connected to an RF clock.
The driving field can also be phase modulated at the clock frequency ($\sim$3.969 GHz). A commercial erbium-doped fibre amplifier (EDFA) is used to amplify the driving signal power. 
The ASE generated in the EDFA is suppressed by a narrow bandpass filter (1\,nm bandwidth 
at 0.5\,dB), centered on the driving laser wavelength. The driving beam is launched into the cavity through the input coupler ($\theta_{\mathrm{in}}$, either 1 or 10\%). 
A polarization controller is used to align the input polarization with one of the two eigenmodes of the cavity.
The cavity transmission is measured by scanning the frequency of the driving laser and recording the average power in the through port with a 200\,kHz photodiode (see e.g. in Figs.\,1B and S2A).

Part of the intracavity power is extracted at the output coupler ($\theta_{\mathrm{out}}$, either 10 or 1\%) to provide a feedback signal for the
cavity stabilization and to characterize the solitons. The cavity detuning is stabilized by slightly changing the driving laser wavelength to maintain a constant intracavity power. The feedback signal is generated by a proportional-integral-derivative controller (Toptica DigiLock 110) driven by the photodiode.
The spectrum of the active cavity soliton (ACS) is recorded on an optical spectrum analyser
(0.1\,nm resolution bandwidth). Time measurements are carried out with a fast photodiode (45\,GHz bandwidth) and an oscilloscope (10\,GHz bandwidth, 10 GSample.s$^{-1}$).
Prior to the photodiode, the signal is filtered by a $\sim$100\,GHz bandpass filter, shifted 2\,nm from the driving laser wavelength, to remove the CW background and increase the SNR of the generated electrical signal~\cite{leo_temporal_2010}.
Alternatively, radio-frequency (RF) measurements and intensity autocorrelation traces are acquired directly at the cavity output. For the autocorrelation measurement, a commercial EDFA is used to increase the average output power to $\sim$100\,mW.

\subparagraph*{\hskip-10pt ACS excitation under CW driving}\ \\
For these experiments, the CW driving beam is injected in the cavity through the 90/10 coupler ($\theta_{\mathrm{in}}=10\%$).
The bias voltage of the amplitude modulator (AM) is set so that, after amplification, the CW driving power reaches $P_s=110\,$mW and a 45\,W flat-top (250 ps duration) single modulation pulse can be added on demand.
The intracavity power is stabilized to approximately 60\,mW, corresponding to a detuning $\delta_0 = 0.5$.
Once the cavity is stabilized, an RF switch connected to the pattern generator is used to add the 45\,W addressing pulse every 5\,ms \cite{wang_addressing_2018}. A single pulse is sufficient to excite or erase ACSs.

\subparagraph*{\hskip-10pt ACS excitation under synchronous driving}\ \\
In this set of experiments, the 99/1 coupler is used as an input port ($\theta_{\mathrm{in}}=1\,\%$). The amplitude and the phase modulators are used to synchronously drive the cavity and to lock the temporal position of the solitons respectively.    
The Turing instability enables the formation of a single soliton without sending an addressing pulse.
We start by stabilizing the cavity on the lower branch of the bistable region, which corresponds to the slope of the backward scan. We choose a PID setpoint of 44~\microw after the output coupler ($\theta_{\mathrm{out}}=10\,\%$), corresponding to a detuning of 0.3 and a quasi-CW background of 83~mW inside the cavity (see black dot in Fig.~\ref{fig_D_250}b). Once in a steady state, we perturb the system by increasing the setpoint, so as to reach the modulationnaly unstable region, and then bring the setpoint back to its original value. This technique is a slower, manual version of the frequency "kick" technique implemented in microresonators
~\cite{volet_micro-resonator_2018} and reproducibly leads to the generation of a single ACS at $\mathcal{D}=1/250$ and multiple ACSs at $\mathcal{D}=1/6$ (see Fig.\,S3b).

\subparagraph*{\hskip-10pt Phase noise measurements}\ \\
For all RF measurements depicted in Fig.~\ref{fig:RF}, the output coupler is the 90/10 ($\theta_{\mathrm{out}}=10\%$). 
The output power is sent to a low noise photodiode (1\,GHz bandwidth) connected to an RF spectrum analyzer with phase noise measurement capabilities (Agilent MXA-9020A). For the soliton state measurement, a $\sim$100\,GHz bandpass filter, shifted 2\,nm from the driving laser wavelength, is added prior to the photodiode. 
The results shown in Fig.~\ref{fig:RF}c have been averaged over five recordings. Each recording takes about 5 minutes.

\section*{Acknowledgements}

\noindent We are grateful to Julien Fatome, Pascal Kockaert, Bart Kuyken and Kasper Van Gasse for fruitful discussions. This work was supported by funding from the European Research Council (ERC) under the European Union’s Horizon 2020 research and innovation programme (grant agreement No 757800). N.E. acknowledges the support of the "Fonds pour la formation à la Recherche dans l’Industrie et dans l’Agriculture" (FRIA, Belgium). P.P.-R. acknowledges the support of the "Fonds de la Recherche Scientifique" (FNRS, Belgium).

\section*{Author Contributions}

\noindent N.E. performed the experiments and simulations of the lumped-element model. N.E. and C.M.A. derived and simulated the mean-field model. C.M.A. and P.P.-R. performed the numerical parameter continuation of the steady state solutions. S.P.G. and F.L. supervised the work.

\section*{Data Availability}
\noindent The data that support the findings of this study are available from the corresponding author upon reasonable request.

\section*{Competing Financial Interests statement}
\noindent N.E., S.P.G. and F.L. filed patent applications on the active resonator design. The remaining authors declare no competing interests.

\bibliography{ACS} 
\bibliographystyle{naturemag}

\clearpage


\section*{Supplementary material}

# Supplementary Information – Temporal Solitons in a Coherently Driven Active Resonator

Nicolas Englebert,[1, *] Simon-Pierre Gorza,[1] and François Leo[1]

[1]*Service OPERA-photonique, Université libre de Bruxelles (U.L.B.),
50 Avenue F. D. Roosevelt, CP 194/5, B-1050 Brussels, Belgium*

This article contains the Supplementary Information for the manuscript entitled "Temporal Solitons in a Coherently Driven Active Resonator". We detail the model used to perform the numerical simulations presented in the article as well as explain how experimental parameters have been extracted. We give more insight about the design of the gain medium. Finally, additional experimental results are discussed.

## Generalized Ikeda map

The evolution of the slow varying envelope of the optical field $u$ is described by the generalized nonlinear Schrödinger equation [1]

$$\frac{\partial u(z,t')}{\partial z} = \left(-\frac{\alpha_0}{2} - \frac{i\beta_2}{2}\frac{\partial^2}{\partial t'^2}\right) u(z,t')$$
$$+ i\gamma \left(1 + \frac{i}{\omega_0}\frac{\partial}{\partial t'}\right) \left(\int_{-\infty}^{\infty} R(t'')|A(z,t'-t'')|^2 dt''\right) u(z,t'), \quad (1)$$

where $z$ is the position along the fibre, $t' = t - \beta_1 z$ with $\beta_1 = [d\beta(\omega)/d\omega]|_{\omega_0}$ where $\beta(\omega)$ is the propagation constant, $\beta_2 = [d^2\beta(\omega)/d\omega^2]|_{\omega_0}$ is the group velocity dispersion, $\alpha_0$ is the fibre loss coefficient, $\omega_0$ is the carrier frequency of the driving field and $\gamma$ is the nonlinear parameter of the fibre. $R(t) = (1-f_R)\delta(t) + f_R h_R(t)$ is the normalized nonlinear response function where $f_R$ is the fractional contribution of the delayed Raman response to the nonlinear polarization and $h_R(t)$ is the Raman response function [1]. The following periodic cavity boundary condition is applied at every roundtrip

$$u^{(m+1)}(0, t') = \sqrt{\rho_{\text{in}}} \, e^{-i\delta_0} u^{(m)}(L, t') + \sqrt{\theta_{\text{in}} P_s} \quad (2)$$

where $\rho_{\text{in}}$ is the input coupler transmission ($\rho_{\text{in}} + \theta_{\text{in}} = 1$), and $\theta_{\text{in}} P_s$ is the driving power injected into the cavity. The cavity detuning is defined as $\delta_0 = 2\pi k - \varphi_0$, where $\varphi_0$ is the overall cavity round-trip phase shift and $k$ is the order of the closest cavity resonance. $L = L_c + L_f$ is the total cavity length.

The insertion losses of the couplers and the isolator are locally applied as

$$u(z_l^+, t') = \sqrt{T_l}\, u(z_l^-, t'), \quad (3)$$

where $l$ denotes the $l^{th}$ fibre component of the cavity, located at position $z_l$, and where $T_l$ is its respective power transmission coefficient.

The gain section is computed in the frequency domain, discretized over the cavity modes of frequency $\nu_{s,j}$, by applying the amplification factor $G(2\pi\nu_{s,j})$ to the input signal and by adding the single pass ASE power $P_A(2\pi\nu_{s,j})$:

$$u(z_{edf}^+, t') = \mathcal{F}^{-1}\left[G(2\pi\nu_{s,j})U(z_{edf}^-, 2\pi\nu_{s,j}) + \sqrt{P_A(2\pi\nu_{s,j})}\,e^{i\phi_j}\right], \quad (4)$$

where $U(z, 2\pi\nu_{s,j})$ is the Fourier discrete transform of $u(z,t')$ and $\phi_j$ is a random phase. $\mathcal{F}^{-1}[\cdot]$ stands for the inverse discrete Fourier transform operator. The bandpass filter is also applied in the frequency domain through its spectral transmission $F(2\pi\nu_{s,j})$:

$$u(z_{bpf}^+, t') = \mathcal{F}^{-1}\left[F(2\pi\nu_{s,j})U(z_{bpf}^-, 2\pi\nu_{s,j})\right]. \quad (5)$$

At 1550 nm, the gain provided by $Er^{3+}$ ions is described by a three-level system. When pumped at 1480 nm, the third level is identical to the second one in the sense that they both belong to the same multiplet. This allows the reduction to a two-level system. We write the total ion density as $N = N_1 + N_2$, where $N_1$ (resp. $N_2$) is the density in the lower (resp. upper) level. In the following, we consider that the signal and the ASE are broadband. The spectrum of the signal [resp. ASE] over one roundtrip is discretized over the cavity modes of frequency $\nu_{s,j}$ with $P_{s,j} = P_s(2\pi\nu_{s,j})$ [resp. $P_{A,j} = P_A(2\pi\nu_{s,j})$]. The population density in the upper level is given by: [2]

$$N_2 = \frac{\sum_j \frac{\tau\sigma_{s,j}^a}{Ah\nu_{s,j}}\Gamma_s P_{s,j} + \sum_j \frac{\tau\sigma_{s,j}^a}{Ah\nu_{s,j}}\Gamma_s P_{A,j} + \frac{\tau\sigma_p^a}{Ah\nu_p}\Gamma_p P_p}{\sum_j \frac{\tau(\sigma_{s,j}^a + \sigma_{s,j}^e)}{Ah\nu_{s,j}}\Gamma_s P_{s,j} + \sum_j \frac{\tau(\sigma_{s,j}^a + \sigma_{s,j}^e)}{Ah\nu_{s,j}}\Gamma_s P_{A,j} + \frac{\tau(\sigma_p^a + \sigma_p^e)}{Ah\nu_p}\Gamma_p P_p + 1} N, \quad (6)$$

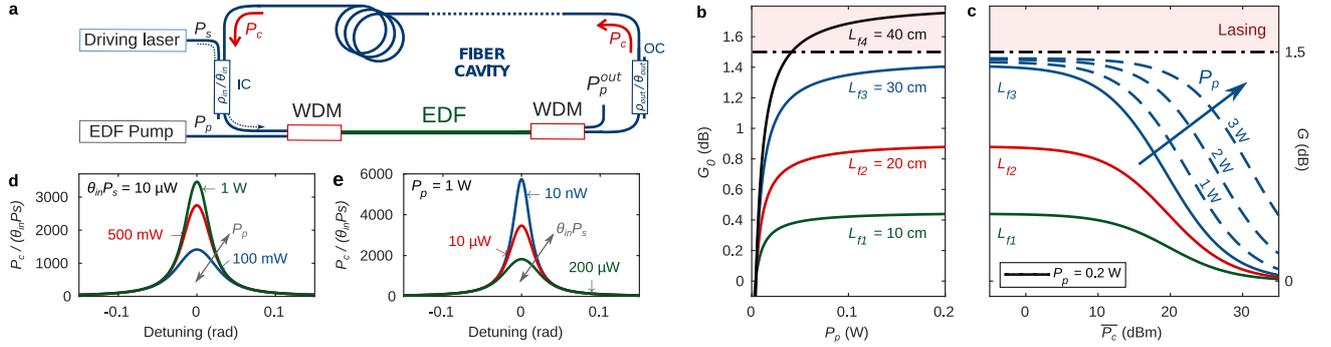

**Fig. S1. Active cavity for ACS generation - Concept and design.** (**a**) Minimalist set-up of the active fibre cavity made of an input coupler (IC), a nonlinear fibre and an amplification section. This latter section is composed of an erbium doped fibre (EDF) and two wavelength division multiplexers (WDMs) to inject / eject the pump power $P_p$. An output coupler (OC) and other components can be added, but they are not strictly required. The cavity is coherently driven by a laser of power $P_s$, leading to a power $P_c$ in the cavity. (**b**) Unsaturated amplification factor $G_0$ as a function of both the pump power and the fibre length $L_f$. The dash-dotted line indicates the lasing threshold. A gain higher than the cavity loss leads to lasing, a situation avoided for ACS generation (red area). (**c**) Amplification factor $G$ as a function of the average intracavity power $\bar{P}_c$. Impact on cavity resonances of the intracavity gain (**d**) and its saturation (**e**).

for a single pump, of power $P_p$ at frequency $\nu_p$, and for a single mode fibre at the pump and signal wavelengths. In this equation, $\tau$ is the lifetime of the upper level, $h$ is the Planck constant and $\sigma_{s,j}^a$, $\sigma_p^a$, $\sigma_{s,j}^e$, $\sigma_p^e$ represent the absorption and emission cross sections at frequencies $\nu_{s,j}$ and $\nu_p$ respectively. $A = \pi R^2$ is the effective area of the erbium distribution with $R$ its radius, and $\Gamma_i$ ($i = s, p$) are the overlap factors between the light-field modes and the erbium distribution. Taking into account that the ASE power is composed of a forward ($+$) and a backward ($-$) wave:

$$P_{A,j} = P_{A,j}^+ + P_{A,j}^-, \quad (7)$$

the propagation of the waves in the EDF is described by the following set of coupled ordinary differential equations: [2]

$$\begin{aligned}
\frac{dP_p}{dz} &= \left(N_2 \sigma_p^e - N_1 \sigma_p^a\right) \Gamma_p P_p - \alpha_0 P_p \\
\frac{dP_{s,j}}{dz} &= \left(N_2 \sigma_{s,j}^e - N_1 \sigma_{s,j}^a\right) \Gamma_s P_{s,j} - \alpha_0 P_{s,j} \\
\frac{dP_{A,j}^+}{dz} &= \left(N_2 \sigma_{s,j}^e - N_1 \sigma_{s,j}^a\right) \Gamma_s P_{A,j}^+ \\
&\quad + 2 N_2 \sigma_{s,j}^e \Gamma_s h \nu_{s,j} \Delta \nu_s - \alpha_0 P_{A,j}^+ \\
\frac{dP_{A,j}^-}{dz} &= -\left(N_2 \sigma_{s,j}^e - N_1 \sigma_{s,j}^a\right) \Gamma_s P_{A,j}^- + \\
&\quad - 2 N_2 \sigma_{s,j}^e \Gamma_s h \nu_{s,j} \Delta \nu_s + \alpha_0 P_{A,j}^-
\end{aligned} \quad (8)$$

where $\alpha_0$ is the intrinsic background loss of the fibre. The solution of these equations gives the amplification factor at each frequency $G(2\pi \nu_{s,j}) = P_{s,j}(L_f)/P_{s,j}(0)$, as well as the ASE spectral density $P_{A,j}^\pm$.

**Numerical simulations**

Numerical integration of (1) over the SMF length $L_c$ is performed through the split-step Fourier scheme [1, 3]. The boundary condition (2) is applied at each roundtrip at the input coupler ($z = 0$). The transmission through the optical isolator is taken into account by (3) before applying (4) for the gain section (see Fig. 1). The amplification factor $G$ and the ASE power $P_A^\pm$ are computed by solving a dual boundary value problem for the system (8) with a relaxation methods over the EDF length $L_f$ [2]. Spectral filtering with (5) directly follows the gain section. The bandpass filter transmission used in our simulations has been recorded experimentally (see right inset of Fig. 3e). Finally, the transmission of the other components (couplers, WDMs, splices, VOA, ...) are applied using (3). The parameters used to simulate the propagation in the SMF are $L_c = 50$ m, with a simulation step of 50 cm, $\beta_2 = -20$ ps$^2$/km, $\gamma = 10^{-3}$ W$^{-1}$m$^{-1}$, $f_R = 0.18$ and $\alpha_0 = 0$. The simulation time window ($2^{19}$ points) corresponds to the roundtrip time (252 ns). The sum of all the loss terms over one round trip is 32 %. The EDF length $L_f$ is 34 cm, and the simulation step to solve (8) is set to 1 cm. The ion density $N = 6.8 \times 10^{24}$ m$^{-3}$ and the cross sections are given by the manufacturer (Liekki®). To reproduce the experimental noise level, an input white noise has been added to the external driving field (0.38 nW in each cavity mode with a random phase), in agreement with the presence of an amplification stage prior to the cavity (see Figs. 3e and S3d). Since the relaxation dynamics of the gain is not taken into account in the model, only the steady state solution has a physical meaning i.e. when the average intracavity power $\bar{P}_c$ reaches its stationary value. Solitons were numerically excited by adding a sech squared pulse to the

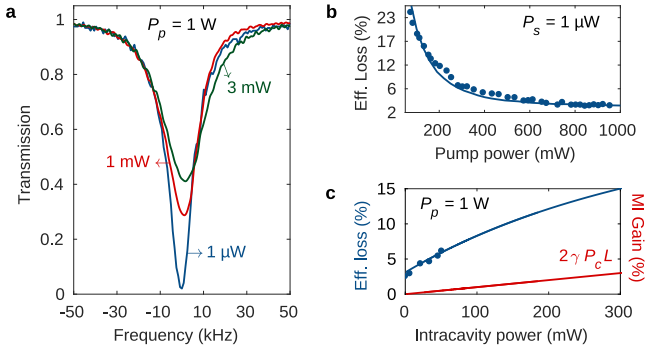

**Fig. S2. Experimental characterization of the cavity transmission**, $\theta_{\rm in} = 1\,\%$. (**a**) Cavity transmission for different input powers: $P_s = 1\,\mu{\rm W}$ (blue), $P_s = 1\,{\rm mW}$ (red) and $P_s = 3\,{\rm mW}$ (green). (**b**) Experimental (dots) and theoretical (line) evolution of the effective loss as a function of the 1480 nm pump power for $P_s = 1\mu{\rm W}$. (**c**) Experimental (blue dots) and theoretical (blue line) evolution of the effective loss as a function of the intracavity power for $P_p = 1\,{\rm W}$. Theoretical modulation instability gain (red line).

steady state intracavity background field [4]. The ACS peak power branch (see Fig. 2a) is obtained by increasing $\delta_0$ in steps of $\Delta\delta_0 = 0.01$.

**Active cavity design**
The coherently driven active fibre resonator considered in this work is a laser cavity pumped below threshold. The purpose of the present section is to give insights into the cavity design for ACS generation.

Figure S1a depicts an active fibre resonator. It is made of an input coupler through which the driving laser is injected, and a gain section. Additional elements such as an output coupler, a dispersive and/or nonlinear section, etc... may also be present. The aim of the gain section is to partially offset the intracavity loss to benefit from low loss resonator properties [see (10)]. For a given EDF, the maximum unsaturated amplification factor $G_0$ is fixed by its length [2]. In Fig. S1b, we plot $G_0$ as a function of the pump power for different fibre lengths, computed by numerical integration of (8) for the Liekki® ER16-8/125 fibre. The gain starts increasing steeply with the pump power and approaches its asymptotic value around $P_p = 200\,{\rm mW}$ for the considered fibre lengths. In the figure, the dash-dotted line represents the cavity loss. Since the goal is to decrease the effective loss $\Lambda_e$ [see (9)] as much as possible, the EDF length should be carefully chosen. For short fibres (e.g. $L_{f1}$ and $L_{f2}$), $\Lambda_e$ always remains high. On the other hand, if the fibre is too long ($L_{f4}$), laser emission occurs, which we seek to avoid. Below threshold, the dynamics is similar to that of a passive resonator in which the coherence of the intracavity field and its amplitude are driven by the injected laser signal ($L_{f3}$ curve). Figure S1c shows the saturation of the amplification factor $G$ for various pump powers $P_p$ when the intracavity power $P_c$ is increased. As can be seen, above a certain value, any increase of the pump power mainly increases the saturation power of the amplifier but barely changes the unsaturated gain [2].

Cavity resonances are directly impacted by the gain and its saturation [5]. The simulated linear resonance for three values of the pump power is shown in Fig. S1d. As expected in the unsaturated regime, increasing the pump power narrows the resonance and the peak power increases, in agreement with a reduced $\Lambda_e$. At a fixed pump power, when the driving power increases, the amplification factor drops from its unsaturated value. Consequently, the resonance broadens and the enhancement factor $P_c/(\theta_{\rm in}P_s)$ decreases. Simulations for $P_p = 1\,{\rm W}$ and various injected powers $\theta_{\rm in}P_s$ are shown in Fig. S1e. Note that at fixed $\Lambda_e$, pumping through a coupler with a large $\theta_{\rm in}$ value decreases the driving pump threshold for ACS generation without impacting the intracavity resonance.

**Experimental resonances and effective loss extraction**
The cavity transmittance, as we increase the CW driving power, for $P_p = 1\,{\rm W}$ is shown in Fig. S2a. The resonances broaden as the intracavity power increases as expected from gain saturation. Fitting the experimental resonance with the theoretical expression of the cavity transmission allows to retrieve the effective loss as a function of the intracavity power. By neglecting both dispersion and nonlinearity in the propagation equations, we can write the cavity transmission as:

$$u^{(m+1)} = \sqrt{\rho_{\rm in}\,\rho_{\rm out}\,T\,G}\,{\rm e}^{(-\alpha_0 L)/2}\,{\rm e}^{i\varphi_0}\,u^{(m)} + \sqrt{\theta_{\rm in}P_s}, \quad (9)$$

where $\rho_{\rm out}$ is the transmission coefficient of the output coupler and $T = \prod_k T_k$ denotes the insertion loss of the other components. After introducing the effective loss:

$$\Lambda_e = -\ln(\rho_{\rm in}) - \ln(\rho_{\rm out}) - \ln(T) + \alpha_0 L - \ln(G), \quad (10)$$

the stationary solution of (9) reads:

$$\frac{P_c}{P_s} = \frac{\theta_{\rm in}}{\left(1 - {\rm e}^{-\Lambda_e/2}\right)^2 + 4{\rm e}^{-\Lambda_e/2}\sin^2\left(\frac{\delta_0}{2}\right)}, \quad (11)$$

where $P_c = |u|^2$. When $\theta_{\rm in} < \Lambda_e$, the transmitted power $P_t = |\sqrt{\rho_{in}P_s} - \sqrt{\theta_{in}}\exp(-i\delta_0)u(L,t')|^2$ displays a minimum for $\delta_0 = 0$. We extract $\Lambda_e$ by fitting $P_t(\delta_0 = 0)$ to the experimental minimum for different driving powers (see Fig. S2a). The experimental detuning $\delta_0$ can then be obtained by comparing the intracavity power, measured at the output coupler, with equation (11). We stress that this is valid when $\gamma L P_c \ll 2\pi$.

Figure S2b shows the evolution of the effective loss as a function of the 1480 nm pump power. Also shown are simulation results obtained with the generalized Ikeda map. We find excellent agreement between theoretical and experimental resonances which validates the model

used to simulate the intracavity amplifier. The effective loss as a function of intracavity power is shown in Fig. S2c along with the theoretical predictions (blue).

To explain the absence of modulation instability in our nonlinear CW pumped experiment (see Fig. 2a), we also plot the modulation instability (MI) gain [6]. The threshold is reached when the gain is larger than the effective loss. We readily see that, because the effective loss depends on the power, MI is not expected for low intracavity powers $P_c$ in our system. The MI threshold (not shown) is $P_c = 2.75$ W, corresponding to $\Lambda_e = 28\,\%$. Our cavity is hence not suited for the observation of extended Turing patterns with a CW driving laser, as the amplification factor is only about 4% at that power.

**Synchronous pumping: Additional measurements**

The results obtained at a duty-cycle $\mathcal{D} = 250$ (Fig. 3) demonstrate the existence of ACS with negligible saturation. When the duty cycle is lowered, a higher average power is required to observe ACSs. An example, for $\mathcal{D} = 6$ is reported in Fig. S3. ACSs are obtained with 40 ns pump pulses and 35.5 mW of average driving power (220 mW peak power), at a detuning $\delta_0 = 0.10$. Fig. S3b shows the signal measured with the 45 GHz photodiode. We can see that many solitons are spontaneously generated on a single driving pulse. These ACS are locked to a regular ∼250 ps temporal grid defined by the frequency of the 3.969 GHz phase modulation. The autocorrelation trace of the ACS pulse train is plotted in Fig. S3c. The peak-to-background ratio is smaller than in Fig. 3d, as expected from the larger effective loss ($\Lambda_e = 7\,\%$). The agreement with simulations is excellent and indicates that 3.3 ps ACSs are excited in the cavity (see inset of Fig. S3c). The experimental and theoretical spectra are shown in Fig. S3d. The presence of fringes in the experimental curve comes from the spectral interference between the multiple solitons. Note that the ASE level is significantly lower than in Fig. 3e because of the stronger gain saturation.

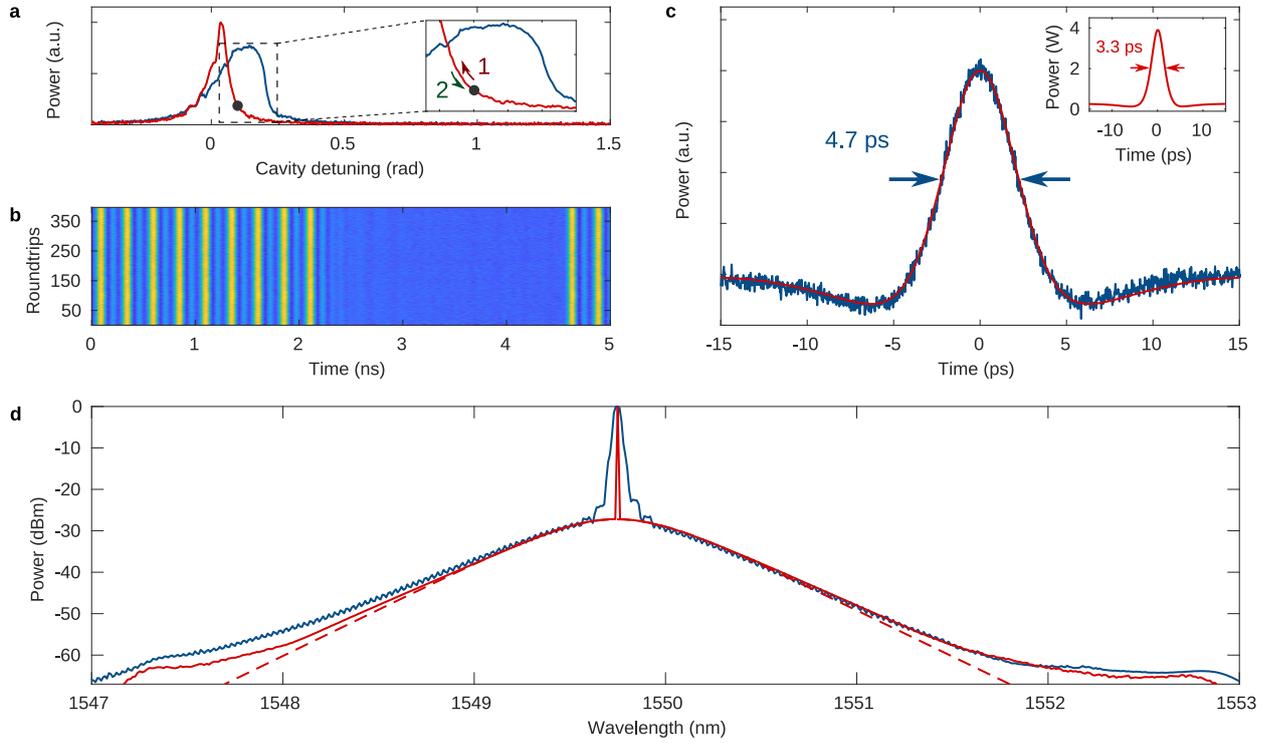

**Fig. S3. Active cavity soliton formation under synchronous pumping** $\theta_1 = 1\%$, $\mathcal{D} = 6$. (**a**) Forward and backward scans through the resonance for $P_s = 220\,\text{mW}$. The grey dot highlights the stabilization setpoint ($\delta_0 = 0.1$). The arrows in the inset illustrate the soliton excitation technique. (**b**) Oscilloscope recording, limited to a 5 ns time window, showing multiple stable short pulses. (**c**) Experimental (blue) and theoretical (red) autocorrelation traces. The inset shows the theoretical temporal profile of the corresponding ACS. (**d**) Experimental (blue) and theoretical (red) spectra measured at the output of the cavity. The dashed red line corresponds to the spectrum of a 3.3 ps wide sech-squared pulse.



\end{document}